# Piezoelectric polyvinylidene fluoride-based epoxy composites produced by combined uniaxial compression and poling

Konstantinos Bezaitis [1], Anthony N. Papathanassiou [1 *] and Elias Sakellis [1,2]

[1] *National and Kapodistrian University of Athens, Physics Department, Solid State Physics Section Panepistimiopolis, GR 15784 Zografos, Athens, Greece*

[2] *National Center of Natural Sciences Demokritos, Institute of Nanomaterials and Nanotechology, Aghia Paraskevi, Athens, Greece*

**Abstract**

We present a simple, efficient method, which combines uniaxial compression and subsequent poling, to produce piezoelectric polyvinylidene fluoride-based epoxy composites. The values of the piezoelectric factors obtained are slightly higher than those of neat piezoelectric polyvinylidene fluoride. The composites respond rapidly and reach a saturation voltage output, to the application of mechanical stimulus quickly. The composites are promising for the creation of *bulk* piezoelectric devices, different from the usual stretched films, exploiting the physic-chemical a of the epoxy matrix. The piezoelectric factor $d_{31}$ versus the mechanical stimulus for the specimens studied, scale according to a double logarithmic representation.

**Keywords:** δ-PVDF; piezoelectric coefficient; relaxation; solid-state-processing; polarization; energy harvesting

(*) Corresponding author; e-mail address antpapa@phys.uoa.gr

Polymer-based composites are being investigated widely due to their ceaseless expansion of application potentials. PolyVinylidene DiFluoride (PVDF) in particular, relishes an overgrowing heed ascribable to its remarkable pyroelectric and piezoelectric attributes [1-10]. Consequently, it serves as a candidate for numerous sensory and actuator devices, mostly in the form of thin stretched $β$-PVDF films [1-6,11]. Manufacturing and optimization of such specimens, often exhibits various levels of complication and scalability issues [6,8-9] which in this work, concurrently with providing longevity and adaptability features, we are attempting to surpass. Issues to be addressed in this work are: (i) the development of a simple and inexpensive experimental scheme for achieving the piezoelectric phase, (b) the efficiency and response of the composite be comparable or better than those of PVDF and composites. (iii) the optimization of the fraction of PVDF, while the composite shares the advantageous physic-chamical features of its matrix (epoxy). The resulting instances will be easily prepared, moldable, durable, efficiently adhesive to different substrates, non-corrosive and remarkably sensitive to mechanical stimuli devices. The protocol followed is uncomplicated compared with that commonly reported for receiving the piezoelectric δ-phase for PVDF microstructure [9]; moreover, it yields bulk specimens of any shape. Furthermore, our experimental configuration permits the determination of an inaccuracy of about 9%, which is better than that reported in published works.

We prepared a test sample of neat PVDF (sample A), an epoxy based PVDF composite (sample B) and another one, loaded with nano-graphene platelets (NGPs) (sample C), the mass ratios of which are given in Table I. Mixtures of non-piezoelectric PVDF powder consist of average molecular weight ~534,000 powder particles with typical size of about 1 μm (Sigma-Aldrich - 182702-100G) and commercial two component epoxy (UHU - 37376) cured at ambient conditions. Moreover, samples with additional nano-graphene platelets (NGPs) (Angstrom - N008-P-40) were prepared as well. The distribution of inclusions are dispersed randomly within the epoxy matrix. The diameter of the resulting disk-shaped specimens are typically a few centimeters and the thickness is typically 1 - 3 mm. For the transition of the as-received PVDF to one of its piezoelectric phases, we followed a two-stage procedure: The specimens were uniaxially compressed under $10^{10}$ $N/m^2$ by using an Enerpack RC 504 hyraulic press for about 5 seconds. The force was applied along axis-1 perpendicular to the parallel surfaces of the composite, while the side free surface area of the sample was un-cofined during deformation; thus, directions perpendicular to axis-1 are indistinguishable and isotropic (axis-3 is labeled any of these axes. The compressed pellets were subsequently polarized within a corona-like schedule; i.e., placed between a couple of parallel spring – loaded platinum electrodes and polarized by an electric field of about 2 $MV/m$ for a time duration of about 5 hours. The advantages of the method applied in the present workare the following: (i) while, in the literature, PVDF is available in films that are stretched to become piezoelectric, in the present work, *bulk* specimens of any size and shape can be formed, and, (ii) the equipment required to attain the piezoelectric phase is simple and inexpensive.

(

To ensure that the above mentioned procedure drives PVDF from the non-piezoelectric α-phase) to its piezoelectric one (based on the literature, it is identical to the δ-phase, induced by axial compression [8]), a solid free standing PVDF pellet (sample A) was formed from its melt and, subsequently, underwent the mechano-electrical scheme described above.

**Table I.** *Mass fraction of epoxy/PVDF composites and NGP.*

| Specimen | (epoxy:PVDF) (w/w) | NGP loading (wt % of the total mass of the composite) |
|---|---|---|
| A | (100:0) | 0 |
| B | (95:5) | 0 |
| C | (95:5) | 1.25 |

The typical dimensions of the disk shaped specimens studied, are a few cm in diameter and about 0.5 mm in thickness. A couple of electrodes permanently silver pasted on the specimen's surface (fig. 1) were connected to a digital Keithley 617 electrometer. Each device was placed inside a Faraday cage for electromagnetic shielding. Real-time voltage or current were recorded as a function of time, while axial force-stimulus covering three orders of magnitude (i.e., $0.02\,N, 0.14\,N, 2\,N$, respectively) could be applied. The sample capacitance was measured directly using Solartron 1260A impedance analyzer controlled by the Novocontrol® WinDETA software.

- *Determination of the piezoelectric coefficient $d_{31}$*

Voltage and current responses to externally applied force for sample B are depicted in figs. 2 and 3, respectively. The force $F_1$ was applied perpendicular to the parallel surfaces of the disk-shaped composite (see fig. 1). Voltage $V_3$ measured between the electrodeis given by:

$$V_3 = \frac{d_{31} \cdot F_1}{C} \tag{1}$$

where $C$ denotes the sample's capacitance and $d_{31}$ is the piezoelectric coefficient. We measured both the response and relaxation for 10 s intervals, for rectangular pulses of mechanical stimuli depicted by dash line in figs. 2 and 3.

The piezoelectric coefficient $d_{31}$ is obtained from eq. (1), whereas $V_3$ holds its saturation value dictated by the application of $F_1$. $d_{31}(F_1)$ data points are plotted in fig. 4, for the three specimens studied in the present work. The determination of the saturation voltage was obdaained from single-pulse stress. However, after thorough experimentation, we led to the conclusion that a time interval between two successive force pulses, such as that depicted by the dash line in figs. 2 and 3, was optimally adequate for the specimen relax compared with the time-scale of the piezoelectric response; the latter is actually immediate to external force triggering. We observe that, for all samples, $d_{31}$ is a decreasing function of $F_1$, in accordance with experimental observations and theoretical modeling for heterogeneous piezoelectric composites. The phenomenon was attributed to the distribution of internal stresses to the piezoelectric inclusions, the local change of the topology and variable degrees of physical contacting of the piezoelectric domains. The piezoelectric coefficient of neat PVDF is comparable or slightly larger than those reported for its piezoelectric phases developed with experimental schemes different than ours (i.e., uniaxial compression and poling). Moreover, epoxy-based composites exhibit $d_{31}$ values compable with those of neat PVDF (specimen A), suggesting that, despite their low content in PVDF (5 wt %), have equal or slightly better efficiency than neat PVDF. The piezoelectric coefficients were were determined with an error than 8 % for samples A and B and 6 – 12 % for sample C. A remarkable feat is the high reproducibility of the results, indicative of a stable structure and predictable behavior of the samples.

The $d_{31}(F_1)$ data points collapse on a single master curve when plotted in a double logarithmic diagram (fig. 5). Regarding the range of mechanical modulus investigated in the present work, a universal empirical fitting function is :

$$d_{31}(F_1) = -5 + 6.136 \times F_1^{-0.178} \tag{2}$$

with correlation factor $R^2 > 0.99$. The description of the experimental results through a common function evidences for a common underlying process for electromechanical coupling and a unified scaled behaiour of the three different composites.

*Determination of the response time to mechanical stimuli*

A functional piezoelectric device is characterized by its ability to requires both strong and rapidly produced voltage output signals. A piezoelectric coefficient is a measure of the ratio of the *saturation* voltage over stress and is a measure of energy conversion. The increase of voltage to saturation upon applying a mechanical stimulus is described by:

$$V(t) \propto \begin{array}{l} 1 - e^{-\frac{t}{\tau}}, \quad F \neq 0 \\ e^{-\frac{t}{\tau}}, F = 0 \end{array} \quad (3)$$

where $\tau$ is a time constant, which tells how fast a saturation voltage value is attained. The values of $\tau$ obtained by fitting eq. (3) to the V(t) data points subsequent to the application of the external force, is presented in fig. 6. At low stress applied, the three specimens share a relatively common short relaxation time (around 2 s) to reach a saturation output. At high stress values, epoxy results in an increase of the relaxation time, while, the small amount of NGPs intensifies the latter phenomenon. Among the three specimens, the one with NGPs (sample C) exhibit a retarded response and differences on exerting or removing the force stimulus. The advantages of the methodology usedm in relation to the current status in the field of piezoelectric sensors and energy harvesting can be summarizes as follows: (i) the experimental scheme for achieving the piezoelectric phase is simple and inexpensive, (ii) the efficiency and response of the composite are improved, compared to those reported , (iii) the composite has optimally low amount of PVDF and shares the advantages of its matrix (epoxy).

In the present work, we developed a simple, reliable and low cost procedure to obtain piezoelectric PVDF and epoxy composites. The piezoelectric $d_{31}$ parameter was determined with high accuracy compared to that usually reported in the literature. Epoxy composites exhibit equal or slightly stronger piezoelectric factors than neat PVDF, probably stemming from their heterogeneous structure. A universal acaling law has been found for $d_{31}(F_1)$, evidencing for a common electro-mechanical coupling mechanism underlying the piezoelectric properties of different samples studied in the present work. Rapid response to applied stimulus is observed.

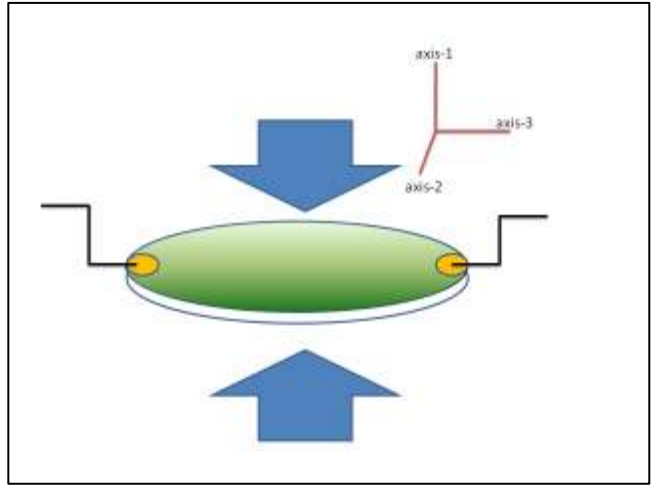

*Figure 1:* *Experimental configuration for measuring the piezoelectric response. Arrows indicate the mechanical stimuli applied along axis-1 and perpendicular to the parallel surfaces of the composite. Electrodes are silver pasted at both ends along axis-3. Leads from the electrodes are connected to a Keithley 617 electrometer (not depicted).*

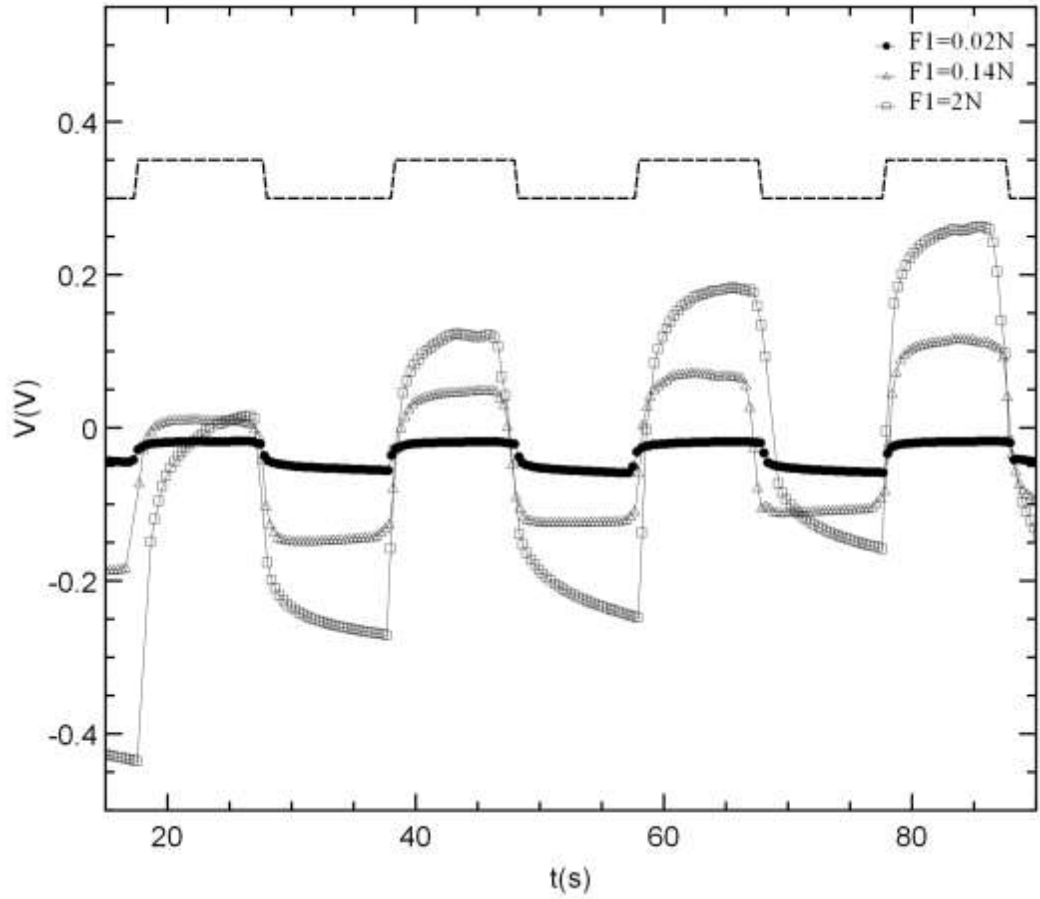

*Figure 21:* *Voltage output measured between the ends of a diameter for a composite with ~5 wt % PVDF (specimen B). Dash line denotes a typical mechanical stimulus (in arbitrary units) as a function of time.*

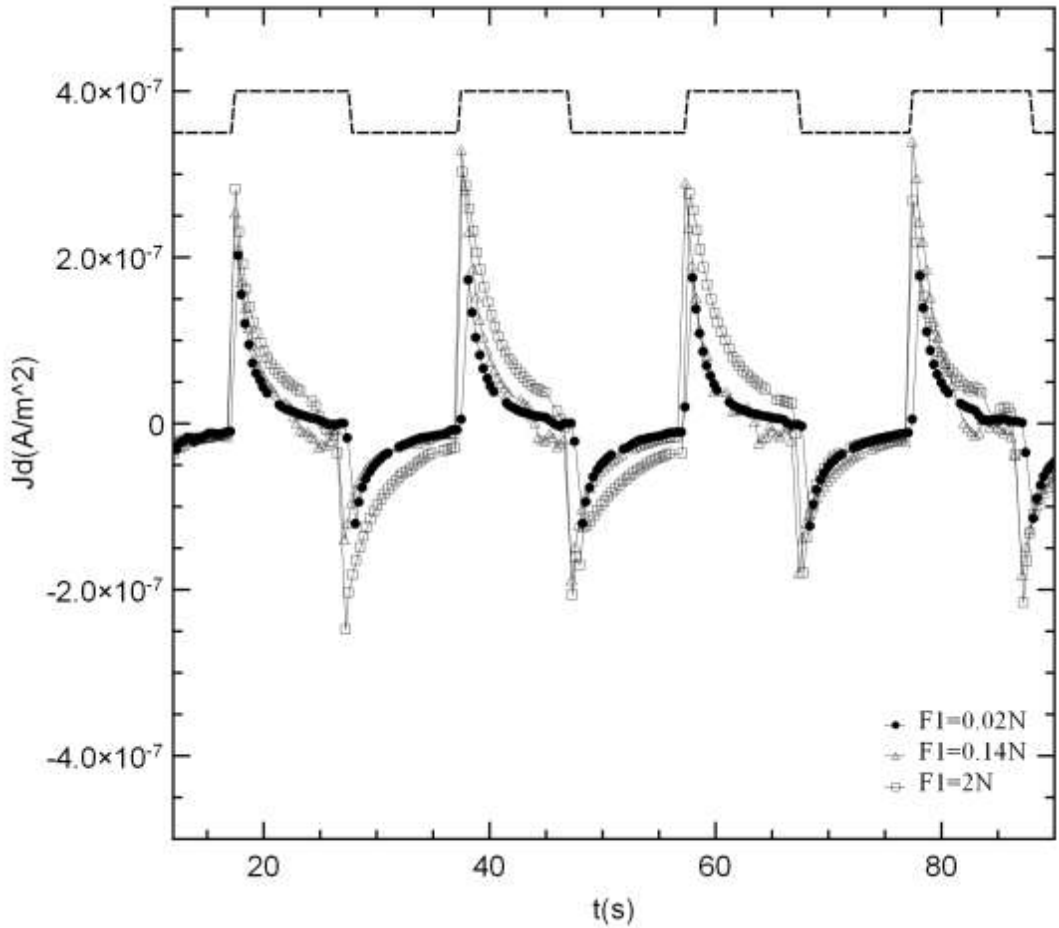

*Figure 3*: *Current density emitted from a composite loaded with 5 wt % PVDF. (specimen B) The sign reversals are signatures of the application and subsequent switch off of the external mechanical stimuli (schematically depicted by the dashed line), respectively. Current evolution is proportional to the rate of the piezoelectric voltage.*

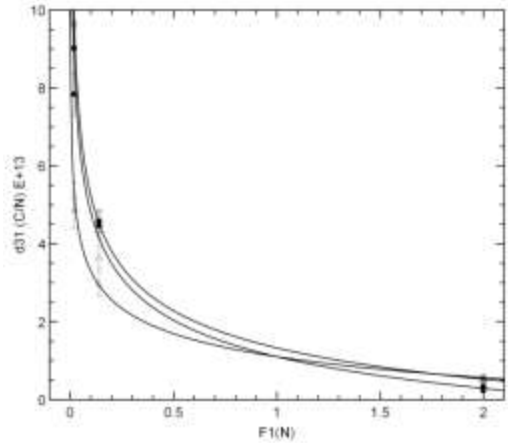

***Figure 4:*** *Piezoelectric coefficients for sample A (circles), B (triangles) and C (squares). Fitting is in the form of* $y = ax^b + c$ *with factor* $R^2 > 0.99$. *The samples provide remarkable performance when 0.02 N is applied and converge while the stimulus strengthens.*

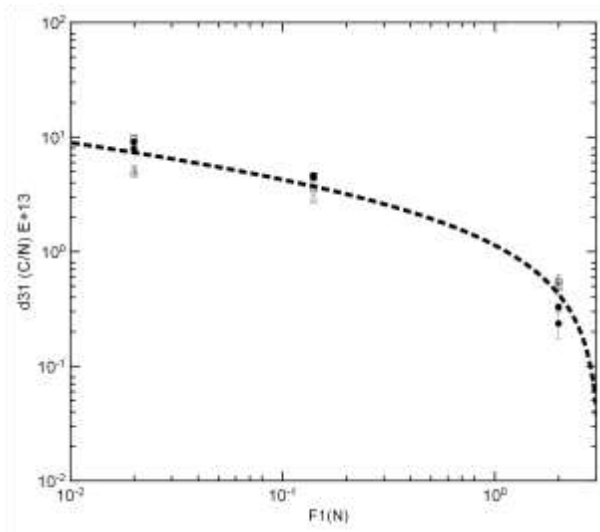

***Figure 5:*** *A master curve (dash line) given by eq. (2) matches the entire set of data points with correlation factor* $R^2 = 0.975$.

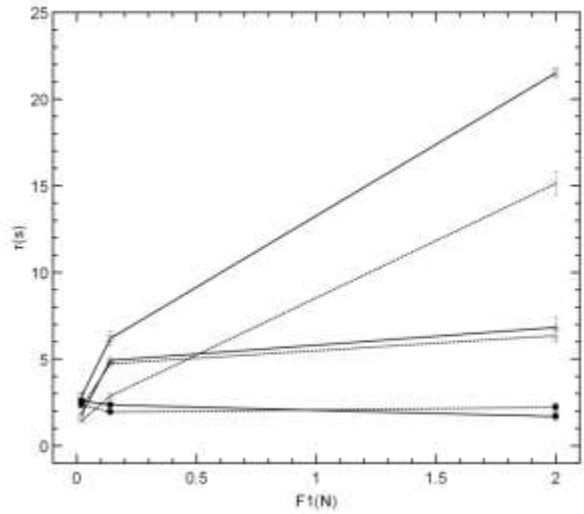

*Figure 6:* Time constants for the response to the application and removal of external mechanical stimulus: sample A (circles), B (triangles) and C (squares). Solid and dashed lines connect data points regarding relaxation after force application and removal, respectively.